# Belief Propagation List Decoding of Polar Codes

Ahmed Elkelesh, Moustafa Ebada, Sebastian Cammerer and Stephan ten Brink

*Abstract*—We propose a belief propagation list (BPL) decoder with comparable performance to the successive cancellation list (SCL) decoder of polar codes, which already achieves the maximum likelihood (ML) bound of polar codes for sufficiently large list size $L$. The proposed decoder is composed of multiple parallel independent belief propagation (BP) decoders based on differently permuted polar code factor graphs. A list of possible transmitted codewords is generated and the one closest to the received vector, in terms of Euclidean distance, is picked. To the best of our knowledge, the proposed BPL decoder provides the best performance of plain polar codes under iterative decoding known so far. The proposed algorithm does not require any changes in the polar code structure itself, rendering the BPL into an alternative to the SCL decoder, equipped with a soft output capability enabling, e.g., iterative detection and decoding to further improve performance. Further benefits are lower decoding latency compared to the SCL decoder and the possibility of high throughput implementations. Additionally, we show that a different selection strategy of frozen bit positions can further enhance the error-rate performance of the proposed decoder.

## I. Introduction

Polar codes [1] were adopted by the 3GPP group as the channel codes for the upcoming *5th generation mobile communication standard (5G)* uplink/downlink control channel [2]. Therefore, decoding of polar codes has been grasping more and more attention as a practical implementation challenge.

As successive cancellation (SC) decoding, originally proposed in [1], is sub-optimal for finite length polar codes, successive cancellation list (SCL) decoding [3] was later introduced achieving the maximum likelihood (ML) bound for a sufficiently large list size $L$, at the cost of increased complexity due to the list decoding nature. Further enhancement of the code was conducted via concatenating a high rate outer code such as Cyclic Redundancy Check (CRC) and parity-check (PC) codes. Under SCL decoding, these CRC-aided polar codes [3] and parity-check concatenated polar codes [4] were shown to outperform the state-of-the-art low-density parity-check (LDPC) codes. Later, an extension of polar codes, namely Polar Subcodes, were proposed in [5], outperforming the above-mentioned code constructions. However, the SCL decoder is characterized by a high complexity and an inherently serial decoding nature, which in turn reduces the decoding throughput and causes high decoding latency (see [6] and references therein). In addition, SCL decoding is not a good match to iterative detection and decoding due to its hard-decision output nature (i.e., not a soft-in/soft-out decoder).

In this work, the major focus is on the achievable performance of iterative decoding of polar codes based on message

The authors are with the Institute of Telecommunications, Pfaffenwaldring 47, University of Stuttgart, 70569 Stuttgart, Germany (e-mail: {elkelesh,ebada,cammerer,tenbrink}@inue.uni-stuttgart.de).
This work has been supported by DFG, Germany, under grant BR 3205/5-1.

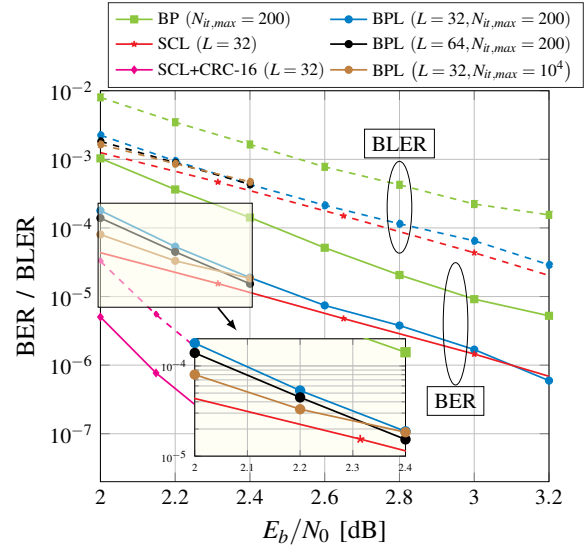

Fig. 1: BER and BLER comparison between BP, SCL ($L=32$), SCL+CRC-16 ($L=32$) and BPL for a $\mathscr{P}(2048, 1024)$-code. All iterative decoders use the **G**-matrix-based stopping condition.

passing over the encoding graph [7]. The BP algorithm enjoys some fundamental advantages over SC-based decoding, as it can be easily parallelized, thus high throughput/low latency implementations are possible, and it inherently enables soft-in/soft-out decoding, facilitating joint iterative detection and decoding. Thus, BP decoding is a promising candidate for high data rate and low latency demanding applications.

Although much effort has been spent on enhancing its performance, e.g., significant improvements through post-processing were reported in [8], the BP decoding algorithm is still outperformed by the SCL decoding algorithm. In [9], it was shown that BP decoding of polar codes can be enhanced by appending a short-length LDPC code to the semi-polarized bit channels of the polar code. In [10], the selection of information and frozen bit channels was altered based on log-likelihood ratio (LLR)-evolution leading to a considerable performance gain. Though these approaches push the performance of BP-based decoding of polar codes a bit forward, the gap between their performance compared to that of SCL decoding is, however, still quite large. Besides, all these methods, making them incompatible to the standardized polar codes (i.e., CRC/PC-aided polar codes).

Based on the observation that the stages of the encoding graph of a polar code of length $N$ can be permuted leading to $(\log_2 N)!$ different graphs with the same encoding behavior, the idea of decoding on parallel BP decoders, each based on a different stage-permuted factor graph realization, was introduced in [11], but has not yet been investigated further. This new decoding scheme was then implemented in a se-quential manner and used in [12] to overcome the effects of insufficient LLR-clipping. It was further studied in [13] where the performance of polar codes under iterative BP decoding with the help of a high rate CRC code was shown to be comparable to SCL (i.e., SCL without the CRC aid). This decoding scheme offers a trade-off between good error-rate performance and large decoding delay (also complexity) due to its sequential decoding fashion. In addition, the decoder approaches the performance of SCL only when the code is concatenated with a high rate CRC code, which is used as a stopping flag in the decoder. Similar permutation decoding techniques were reported in [14] for Reed–Muller (RM) codes.

Unlike [13], where the code itself was changed via concatenating the polar code with a high rate CRC code, the proposed algorithm in this paper proposes a new decoder based on a list of $L$ parallel independent BP decoders without altering the code itself (i.e., no code concatenation is involved), hence it is referred to as "BPL decoder". Compared to the decoder proposed in [13], BPL allows the parallel execution of all graphs and, thus, possibly both higher throughput and lower latency can be achieved. It is worth mentioning that this decoder resembles the one proposed in [15] where Bose-Chaudhuri-Hocquenghem (BCH) and Golay codes are decoded based on multiple BP decoders over different graph realizations. To the best of our knowledge, the proposed BPL decoder provides the best, so far, known performance of plain polar codes under iterative decoding. The decoder is introduced in its basic high complexity (but conceptually simple) form. Further complexity reduction approaches are reserved to future investigation.

## II. POLAR CODES AND ITERATIVE DECODING

As introduced in [1], polar codes are based on the concept of channel polarization, where $N = 2^n$ identical channels are combined such that a channel polarization effect occurs. The recursive channel combination leads to the generator matrix $\mathbf{G}$, given by $\mathbf{G} = \mathbf{F}^{\otimes n}$, where $\mathbf{F} = \begin{bmatrix} 1 & 0 \\ 1 & 1 \end{bmatrix}$ and $\mathbf{F}^{\otimes n}$ denotes the $n$-th Kronecker power of $\mathbf{F}$.

The codewords $\mathbf{x}$ can be obtained by $\mathbf{x} = \mathbf{u} \cdot \mathbf{G}$, where $\mathbf{u}$ contains $k$ information bits and $N - k$ frozen bits. The $k$ information bits are assigned to the $k$ most reliable positions denoted as information set $\mathbb{A}$ (and $\bar{\mathbb{A}}$ denotes the frozen positions), which have to be found during the code construction phase. If not explicitly mentioned, we use Arıkan's Bhattacharyya-based design; however, any other polar code construction (i.e., different $\mathbb{A}$) could be used. Let $\mathscr{P}(N,k)$ denote a polar code with codeword length $N$ and $k$ information bits, i.e., the information set has the cardinality $k = |\mathbb{A}|$.

An iterative decoding algorithm was introduced in [7] based on Gallager's BP decoding of LDPC codes. The main difference is that the iterative message passing is applied over the encoding factor graph (Fig. 2) instead of the parity-check matrix. This factor graph consists of $n+1$ stages containing $N$ nodes per stage. Due to the limited space, only a short overview on the BP decoding algorithm is provided. We refer the interested reader to [7], [12] and [16]. Two types of messages are involved, left-to-right messages (**R**-messages) and right-to-left messages (**L**-messages). The **R**-messages at stage 1 represent the a priori information available to the decoder and thus, are either 0 or $\infty$ for non-frozen ($i \in \mathbb{A}$) and frozen ($i \in \bar{\mathbb{A}}$) bits, respectively. The **L**-messages at the last stage $(n+1)$ are initialized with the channel output LLRs $L_{ch,i} = \log \frac{P(x_i=0|y_i)}{P(x_i=1|y_i)}$, i.e., $L_{i,n+1} = L_{ch,i}$.

During each iteration, **L**- and **R**-messages are iteratively propagated from stage to stage through the factor graph [7]. As shown in Fig. 2, the polar factor graph consists of $\log_2(N) \cdot \frac{N}{2}$ processing elements (PE)s (or check nodes, CNs) [12].

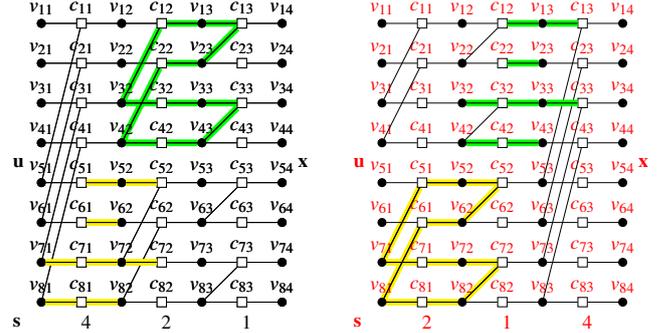

(a) Conventional factor graph ($\mathbf{s} = [4,2,1]$)  (b) Stage-shuffled factor graph ($\mathbf{s} = [2,1,4]$)

Fig. 2: Factor graphs of $\mathscr{P}(8,k)$-code (loop structure changes).

The BP decoder can stop after a pre-defined maximum number of iterations $N_{it,max}$, or if an early stopping condition is fulfilled [16]. Finally, either the LLRs of the estimated information word $\hat{\mathbf{u}}$ and the estimated codeword $\hat{\mathbf{x}}$ is forwarded as the decoder output, or a hard decision is applied to recover the information bits. Two stopping conditions, based on intermediate hard-decisions on $\hat{\mathbf{u}}$ and $\hat{\mathbf{x}}$, are practically relevant:

1) **G**-matrix-based [16]: stop when $\hat{\mathbf{x}} = \hat{\mathbf{u}} \cdot \mathbf{G}$ is fulfilled.
2) CRC-based: stop when the CRC on $\hat{\mathbf{u}}$ is fulfilled. However, this would require an outer CRC code (i.e., a change in the code structure).

## III. BELIEF PROPAGATION LIST DECODER

As outlined in [11], there exist $n!$ different permutations of the polar code factor graph which fulfil the condition $\mathbf{x} = \mathbf{u} \cdot \mathbf{G}$. Any permutation of the factor graph based on Arıkan's kernel can be characterized by a vector $\mathbf{\Pi}$ with entries $\pi_j$ from 1 to $n$ permuted in any random order, written as: $(\mathbf{\Pi})_{1 \times n} = \text{randperm}(n)$ (i.e., random permutations of the vector $[1,2,3,\ldots,n]$). The $j^{th}$ stage in the factor graph is characterized by $s_j$ defining the separation between the two bit channels to be connected together via a check node. The vector $\mathbf{s}$ with entries $s_j$ for the $n$ stages is calculated as: $s_j = 2^{\pi_j - 1}$, where $1 \leq j \leq n$. Fig. 2a and 2b show a pair of permutations for $\mathscr{P}(8,k)$-code (i.e., $n=3$). The factor graph in Fig. 2a is defined by the vector $\mathbf{\Pi} = [3,2,1]$ and thus the separation $\mathbf{s} = [4,2,1]$. While the factor graph in Fig. 2b is defined by the vector $\mathbf{\Pi} = [2,1,3]$ and thus the separation $\mathbf{s} = [2,1,4]$[1].

---

[1] Remark: the decoder can be sufficiently described by $\mathbf{s}$, however, from an implementation perspective we believe that a permutation $\mathbf{\Pi}$ of the natural numbers from 1 to $n$ is more convincing.

BP decoding can be performed over any one of such permuted factor graphs. Different permutations contain different loops as depicted in Fig. 2. Thus, one factor graph permutation may be better than another permutation for a specific set of inputs (i.e., transmitted codeword plus noise realization).

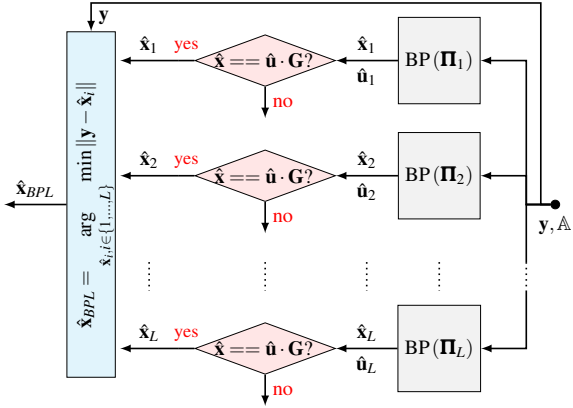

Fig. 3: An abstract view of BPL decoding ($L$ parallel independent BP decoders).

*A. Algorithm*

As there exist no clear evidence on which graph permutation performs best for a given input, the BPL decoder of polar codes consists of $L$ parallel independent BP decoders[2] each based on a different permuted factor graph. Some cyclic shifts of the original factor graph (i.e., the vectors $\Pi_k$ are cyclic shifts of $[1, 2, 3, \ldots, n]$, where $1 \leq k \leq L$) were observed to be better than randomly permuted factor graphs. Thus, and for reproducibility, the $n-1$ cyclic shifts were always selected among the $L$ different factor graph permutations. The full decoding procedure is depicted in Fig. 3. BP iterations are conducted on each single decoder until the **G**-matrix-based stopping condition is satisfied, or a maximum number of iterations per decoder $N_{it,max}$ is performed. Therefore, after $N_{it,max}$ iterations, a list of $L$ codewords (i.e., candidates) is available. Only those codewords satisfying the **G**-matrix-based stopping condition are declared as the set of valid polar codewords. Finally, that valid codeword from the $L$ parallel BP decoders which is closest, in terms of Euclidean distance, to the channel output $\mathbf{y}$ is picked to be the BPL decoder output

$$\hat{\mathbf{x}}_{BPL} = \arg\min_{\hat{\mathbf{x}}_i, i \in \{1,\ldots,L\}} \|\mathbf{y} - \hat{\mathbf{x}}_i\|.$$

Fig. 1 shows that the error-rate performance of the BPL decoder converges to that of the SCL decoder for the same information set $\mathbb{A}$. One can also infer from Fig. 1 that increasing $L$ and $N_{it,max}$ enhances the BPL error-rate performance.

*B. Potential Benefits*

The BPL is a soft-output decoding algorithm and, thus, can be used in joint detection and decoding which is not possible with hard-output decoders (e.g., SCL).

---

[2]The major difference between SCL and BPL decoders in building the list is that the SCL is based on the branching-while-decoding strategy; thus, the list size is doubled after a bit decision until reaching $L$ branches, while the BPL decoder starts with $L$ branches (i.e., $L$ parallel independent BP decoders).

As shown in [6], iterative decoding of polar codes has a higher potential of high throughput implementations when compared to SC-based decoding. Thus, BPL enjoys a higher potential of parallel hardware implementations leading to high throughput and low latency. The latency of BPL scales with $\mathcal{O}(N_{it,max} \cdot \log N)$ as each stage can be fully parallelized, i.e., it scales with $\log_2 N$ for a constant number of iterations. Furthermore, stage-combining is possible without any performance loss, reducing the decoding latency and the memory complexity. In contrast, due to the sequential decoding nature of SCL, low latency implementations are hard to realize (cf. [6]), i.e., the latency of SCL scales with $\mathcal{O}(N)$ without further pruning techniques such as partitioning.

A major advantage of our proposed decoder, namely BPL, is the reduced sorting complexity when compared to the SCL decoder. In SCL decoding, for large values of $L$, the metric sorter is time consuming and expensive in terms of complexity, while in BPL decoding no intermediate sorting steps are needed; only one sorting operation is done at the very end to pick the closest codeword to $\mathbf{y}$.

The price, however, to pay is having $L$ parallel BP decoders with an overall complexity of $\mathcal{O}(L \cdot N \cdot \log N)$ assuming a constant number of iterations. It is worth mentioning that the complexity of BPL increases linearly with the number of performed BP iterations. Thus, in terms of the number of performed operations, a naive implementation of BPL is more computationally complex than SCL. A basic SCL ($L = 32$) implementation for a $\mathscr{P}(2048, 1024)$-code requires $0.7209 \times 10^6$ unit calculation updates [17]. BPL ($L = 32, N_{it,max} = 200$) requires $25 \times 10^6$ PE updates at $\frac{E_b}{N_0} = 2$ dB and $14 \times 10^6$ PE updates at $\frac{E_b}{N_0} = 3.2$ dB, respectively. However, in the high SNR region (e.g., $\frac{E_b}{N_0} = 3.2$ dB), BPL only requires a list size $L = 5$ to achieve the SCL ($L = 32$) error-rate performance and, thus, requires $1.01 \times 10^6$ PE updates. PE and unit calculation updates have similar complexity. Thus, in the high SNR region and without further complexity reductions, the BPL has a feasible complexity when compared to plain SCL.

## IV. RM-POLAR CODES UNDER BPL DECODING

We observed that the BPL decoder benefits only slightly from an outer CRC/PC code. However, we show in this section that a more proper polar code design can further reduce the gap to the CRC-aided SCL performance by simply selecting the frozen/information bit positions according to [18].

On the one hand, RM and polar codes are based on the same polarization matrix $\mathbf{F}^{\otimes n}$. On the other hand, they differ in the selection of bit channels used for conveying the information bits. In polar codes, the information bits are those $k$ bits with the $k$ smallest Bhattacharyya parameters; while in RM codes, the information bits are the $k$ bits corresponding to the highest row weights (i.e., Hamming weights) in the generator matrix $\mathbf{G}$. The goal of the polar code construction is to minimize the error probability under SC decoding, which is not necessarily optimum under iterative decoding (i.e., "flooding" BP decoding). However, the goal of the RM construction is to design a code with the largest minimum Hamming distance. For finite length codes, the distance properties are of more significant impact than the polarization effect [19].

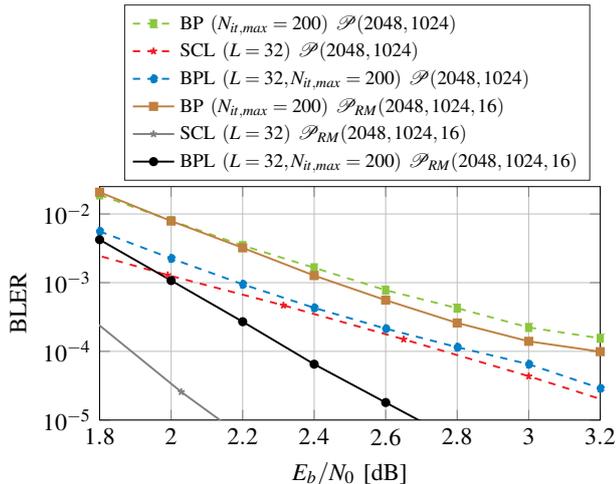

Fig. 4: BLER comparison between BP, SCL ($L=32$) and BPL $\left(L=32, N_{it,max}=200\right)$ for both $\mathscr{P}(2048,1024)$-code and $\mathscr{P}_{RM}(2048,1024,16)$-code. All iterative decoders use the **G**-matrix-based stopping condition.

In [18], a new type of codes is constructed according to both the RM and polar code construction rules together, yielding a code with a higher minimum Hamming distance than polar codes, and thus, better error-rate performance in the finite length regime. First, the bits corresponding to a row in the **G**-matrix which has a weight less than or equal to a certain threshold $d$ are set to be frozen, regardless of the Bhattacharyya parameter values of these bit channels. Afterwards, the $k$ information bits are the ones with the smallest Bhattacharyya parameters from the remaining set (i.e., the most reliable $k$ bit channels under SC decoding). Finally, the remaining set of bits are set to be frozen. An RM-polar code of length $N$ with $k$ information bits and the Hamming weight threshold $d$ is denoted by $\mathscr{P}_{RM}(N,k,d)$.

This hybrid RM-polar code construction method yields codes that have significantly better error-rate performance than the codes constructed by the Bhattacharyya parameter under BPL and SCL decoding, Fig. 4. Note that an SCL decoder of a $\mathscr{P}_{RM}(2048,1024,16)$-code requires a pretty large list size $L$ ($L > 512$) to closely approach the ML bound of the hybrid RM-polar code [18]. For the case of the $\mathscr{P}(2048,1024)$-code under SCL decoding, it was shown in [3] that a list size of 32 is sufficient to closely approach the ML bound of the polar code in the high SNR region ($E_b/N_0 > 1.5$ dB).

In Fig. 4 it can also be seen that the hybrid RM-polar code construction method yields codes that have better error-rate performance than the codes constructed by the Bhattacharyya parameter under BPL decoding. This gain comes at no extra complexity cost, where the only difference between the codes is the frozen/information bits selection. It should be emphasized that no additional CRC was used for the results in Fig. 4.

## V. CONCLUSION

We propose a multiple permuted factor graph-based iterative decoding algorithm of polar codes which reaches the same error-rate performance as SCL decoding. To the best of our knowledge, this BPL decoder is the best iterative decoder for plain polar codes known thus far. The proposed algorithm does *not* require any changes or additions in the polar code structure, i.e., the gains are solely based on the improved decoding algorithm itself. Additionally, a hybrid RM-polar code design can further significantly enhance the performance as it improves the distance spectrum of the polar code. We show results for carefully constructed polar codes under iterative (i.e., BPL) decoding, surprisingly only less than 0.5 dB away from the CRC-aided polar codes under SCL decoding. The advantages of the proposed decoder are soft-in/soft-out decoding and the possibility of low latency implementations. The drawbacks, high decoding complexity and the missing CRC/PC compatibility, are left open for future research.